# The Group Administered Interactive Questionnaire: An Alternative to Individual Interviews

Edit Yerushalmi[1], Charles Henderson[2], William Mamudi[2], Chandralekha Singh[3], Shih-Yin Lin[3]

[1]Department of Science Teaching, Weizmann Institute of Science, Rehovot 76100, Israel
[2]Department of Physics and Mallinson Institute for Science Education, Western Michigan University, Kalamazoo, MI 49008, USA
[3]Department of Physics and Astronomy, University of Pittsburgh, Pittsburgh, PA 15213, USA

**Abstract.** Individual interviews are often considered to be the gold standard for researchers to understand how people think about phenomena. However, conducting and analyzing interviews is very time consuming. This paper presents the Group Administered Interactive Questionnaire (GAIQ) as an alternative to individual interviews and discusses the pros and cons of each data collection method. Use of GAIQ will be discussed in the context of a study that seeks to understand teaching assistants' reasons for the design of problem solutions for introductory physics.

## INTRODUCTION

A former line of research [1] investigated physics faculty beliefs and values about the teaching and learning of problem solving. Our current work builds on this former line of research to investigate graduate teaching assistants' beliefs about the role that worked examples should play in introductory physics instruction. Graduate teaching assistants play a central role in the teaching of problem solving. Teaching assistants lead recitations in which they present students with worked-out examples of physics problems, guide students in solving problems and assess students' solutions.

The data collection for the study of faculty beliefs made use of individual semi-structured interviews that took an "artifact comparison" approach. Namely, interviewees were shown classroom artifacts (i.e. instructor solutions) and were asked pre-defined questions (i.e. "describe how these instructor solutions are similar/ different to your own practice, explain your reasons"). The data collected in this manner allowed for a comprehensive description of faculty beliefs [1]. However, there are several concerns regarding this method. First, from the practical perspective, it requires significant time for both data collection and analysis. Second, the interviewer interventions required to clarify respondents' answers, endanger reliability. Finally, as the data collected is extremely rich, there is ambiguity in categorization of the data, endangering validity.

In this paper, we present an alternative that we have developed, the Group Administered Interactive Questionnaire (GAIQ), and used to investigate considerations that shape teaching assistants instructional choices regarding worked-out examples of physics problems. The GAIQ was designed to respond to the aforementioned concerns. On one hand, it aimed to maintain the process of clarifying respondent's ideas which is a key feature of the interview method. On the other hand, it aimed to require less time for data collection and analysis, avoid researcher intervention in the process of clarifying the respondents' answers and allow the respondent to assist in the categorization. We will first describe the GAIQ method and then reflect on how it compares to in-person interviews.

## THE GAIQ METHOD

In our interview study, interviewee's considerations about the design of instructor solutions are clarified through discussion between the interviewer and the interviewee. GAIQ took advantage of a methods course for physics graduate assistants at University of Pittsburgh. Twenty-four TAs participated in this one semester weekly course. The GAIQ replaced the one-on-one discussion that takes place in an interview with a sequence of activities that took place during the first three weeks of the course. Table 1 summarizes this sequence.

**TABLE 1:** GAIQ sequence of activities.

| Time | Activity |
|------|----------|
| Pre | Individually, students solve target problem (Fig 1) and answer **questions** in pre-discussion worksheet that are related to the 3 instructor solutions (Table 2). |
| Lesson | In groups of 3, students answer the **same questions** in group worksheets then a whole class discussion takes place where groups share their work. |
| Post | Individually, students answer **same questions** in post-discussion worksheet (Table 3). |

You are whirling a stone tied to the end of a string around in a vertical circle having a radius of 65 cm. You wish to whirl the stone fast enough so that when it is released at the point where the stone is moving directly upward it will rise to a maximum height of 23 meters above the lowest point in the circle. In order to do this, what force will you have to exert on the string when the stone passes through its lowest point one-quarter turn before release? Assume that by the time that you have gotten the stone going and it makes its final turn around the circle, you are holding the end of the string at a fixed position. Assume also that air resistance can be neglected. The stone weighs 18 N.

**FIGURE 1.** Problem used in study.

In the pre-lesson stage, as part of their homework, TAs were asked to write a problem solution that they would hand out to their students. The problem (Figure 1) was selected to be one that, although difficult, could reasonably be given in most calculus-based introductory physics courses. TAs were then asked to respond to open-ended questions about how they think solved problems should be used in instruction (e.g., "In what situations do you believe students should be provided with examples of solved problems?", "How would you like your students to use the solved examples you give them?" "What do you think most of them actually do?"). The TAs were also provided with three instructor solutions for the problem (two are shown in Figure 2) and were asked to fill in a Pre-discussion Individual worksheet (Table 2) where they identified prominent features of the solutions, ranked the solutions based on i) which solution has more of each feature and ii) their preference for including each features in solutions, and explained their reasons.

**TABLE 2.** Pre-discussion worksheet

| Which features of these solutions would you like to include in solutions you are writing for your students? Please explain your reasons | | | | | | | |
|---|---|---|---|---|---|---|---|
| **Solution features** | **Representation:** Rank the solutions based on which solution has more of this feature. (You could also Mark + for the solutions in which this feature exists.) | | | Rank the solutions based on your **preference** for this feature (A - for the one you like the most in how it represents this feature to C-for the one you like the least) | | | **Reason**: Why do you like/ not like this feature? |
| | Sol. I | Sol. II | Sol. III | Sol. I | Sol. II | Sol. III | |
| | | | | | | | |

**FIGURE 2.** Instructor Solutions I & III.

In the lesson stage the TAs interacted in small groups to share their ideas regarding the use of example problem solutions.

Finally, TAs filled a Post-discussion Individual worksheet (Table 3) where they were provided with the opportunity to modify their previous answers. In addition to being able to modify their previous answers, TAs were asked to connect their instructor solution features to a list of pre-defined features developed by the researchers based on a pilot study (Table 4).

**TABLE 3.** Post-discussion worksheet

| Write down the features' numbers that you originally noticed (use attached feature list (Table 4)). For each of your features, write down how you originally named this feature. Describe how and why, if at all, your preference towards it changed |||||
|---|---|---|---|---|
| Original feature name | Feature number | Rate the solutions based on your **current** preference for this feature: … ||| **In case** your preference towards it changed following the class discussion, elaborate your final preferences: Why do you like or dislike this feature? |
| | | Sol. I | Sol.II | Sol. III | |
| | | | | | |

**TABLE 4.** Pre-defined feature list.

| | |
|---|---|
| 1. | Provides a schematic visualization of the problem (a diagram) |
| 2. | Provides a list of knowns/unknowns |
| 3. | Provides a "separate" overview of how the problem will be tackled (explains premise and concepts -- big picture -- prior to presenting solution details) |
| 4. | Explicit sub-problems are identified (Explicitly identifies intermediate variables and procedures to solve for them) |
| 5. | Reasoning is explained in explicit words (Description/justification of why principles and/or sub-problems are appropriate/useful in this situation) |
| 6. | The principles/concepts used are explicitly written using words and/or basic mathematical representations (e.g., F=ma or Newton's $2^{nd}$ Law) |
| 7. | Thorough derivation (Detailed/verbose vs. Concise/short/simplified/skips lots of derivation ) |
| 8. | Long physical length (Long/verbose vs. Short/concise vs. Balanced/not too long, not to short ) |
| 9. | Includes details that are not necessary for explaining the problem solution (the solution is technically correct and complete without these 'unnecessary' details) |
| 10. | Provides alternative approach |
| 11. | Solution is presented in an organized and clear manner |
| 12. | Direction for the progress of the solution progress: Forward vs. Backward |
| 13. | Symbolic solution (numbers are plugged-in only at the end) |
| 14. | Provides a check of the final result |

The instructor solutions that respondents were asked to make judgments about were carefully designed to activate, in an imaginary classroom setting, the instructional decision-making that takes place in an authentic classroom. Through making and justifying instructional decisions, research subjects expose the beliefs and values that underlie these decisions. The Instructor solutions reflect various instructional styles. None of the solutions were designed to be flawless. For example, Instructor solution I (Figure 2) was a "bare-bones" solution that left many of the minor steps to be filled in by the reader and used a rough sketch. Instructor Solution III is designed to reflect a systematic decision making process characteristic of expert problem solvers. It begins with a detailed diagram and separate overview that states the problem goal and attempts to relate it to the known information while explaining the reasoning behind each step.

## THE GAIQ vs. THE INTERVIEW

Figures 3 and 4 show correspondingly samples of data collected in the interview and in the GAIQ method.

**Instructor:**…*I think it is a little more than that, yeah. I think you've basically…I kind of like this one a little better. You know, he's saying, "I need the force," right. For a massless string the force is equal to the tension on the bottom. He's sort of worried about the weight of the string, and that's something, you know. I sometimes don't even worry about that. "I can relate the force using centripetal acceleration," ok that's good. "I can relate v-sub-b to v-sub-t either using energy," ok. Basically he's talking about how he can do that and, you know, you've got kind of a problem with angles and stuff if you want to do it from mechanics and he doesn't want to do that. You know, so energy is what you want to use on that. I wouldn't do that type of problem without energy in that course.*
**Interviewer:** And you would also tell them, you know, these possibilities?
**Instructor:** *Oh yeah. I try to do that. I try to outline sort of where I'm going to go before I actually get into it. Rather than just launch into the thing and just whip it out. I want to let them know essentially what I'm thinking about and where I'm going, and then I try to solve it…*

**FIGURE 3.** Sample data from Interview.

| TAs filled in pre-discussion worksheet |||||||
|---|---|---|---|---|---|---|
| Feature | Representation ||| Preference (A – best) ||| Reason |
| | I | II | III | I | II | III | |
| *Known/ unknowns* | | | | | $\mathcal{A}$ | | *By listing what you know it's easier to know what you don't know* |
| *approach/ execution* | | | + | $C$ | $B$ | $\mathcal{A}$ | *Knowing the approach – you can apply this to other problems* |

| TAs filled in post-discussion worksheet ||||||
|---|---|---|---|---|---|
| Original feature name | Feature number | Current preference ||| Reason in case of change |
| | | I | II | III | |
| *Known/ unknowns* | 2 | | $\mathcal{A}$ | | *By listing what you know it's easier to know what you don't know* |
| *approach/ execution* | 3 (Provides a "separate" overview) | $C$ | $B$ | $\mathcal{A}$ | *Allows a conceptual idea of progress instead of just plug-chug* |

**FIGURE 4.** Sample data from GAIQ.

One can see that the resulting data in the GAIQ is provided in a more compact manner. In the interview data (Figure 3), it is quite difficult to determine to which feature the instructor is referring. In this case, we believe the respondent referred to feature 3 (Providing a "separate" overview in Table 4), that he perceives this feature to be represented in solution III, and that he likes it. Yet, one could relate the instructor statement also to other features, such as feature 6 (The principles/concepts used are explicitly written). Moreover, the interviewer's questions seem to refer to feature 10 (Providing alternative approach), yet the instructor responds to something else. The intent of the interviewee can be masked in the conversation. In comparison, in the GAIQ data (Figure 4) the TA himself relates his original feature "approach/execution" to feature 3 from the list and defines which solutions represent it best and his preference towards it (A = positive). Thus, GAIQ shifts some of the analysis work from the researcher to the respondent as the respondents themselves differentiate between the features, preferences and reasons and categorize the features.

In GAIQ, the clarification process is done separately and the resulting data includes the initial ideas the TA holds "Knowing the approach – you can apply this to other problems" and their refinement "Allows a conceptual idea of progress instead of just plug-chug".

Table 5 summarizes the comparison between the GAIQ and the interview, both in the data collection process, as well as in the resulting data.

# SUMMARY

In this paper, we have introduced our rationale and methods for uncovering instructors' considerations in presenting students with worked-out examples for physics problems. The design of our data collection tool accounted for reliability via pre-determined questions focused on concrete instructional artifacts that instructors routinely design and assess. This served to standardize data collection in order to collect reproducible data in both the interview and the GAIQ.

As suggested by Kvale [2], the validity of data gathered through an interview is the result of an inter-subjective agreement that an interviewee and interviewer reach in a process of clarifying meaning. With this conception of validity in mind, we replaced the process of clarifying meaning via interviewer probing questions with a group discussion aimed at sharing and articulating participants' ideas till a mutual understanding has been reached. Thus, the GAIQ data collection tool allowed us to arrive at knowledge of how people think about the same phenomenon probed by the interview, maintaining to some extent the clarification process, while avoiding intervention of an interviewer. A related paper [3] presents analysis of the resulting data to determine TAs' considerations in presenting students with worked-out examples.

**TABLE 5.** Comparison: GAIQ vs. the interview

| Dimension | Interview | GAIQ |
|---|---|---|
| Duration | Travel time + F2F interviews with all respondents | One lesson + photocopying TAs' worksheets |
| Time for respondents to think | Respondents have little time to reflect on their answer | Respondents have time to reflect on their answers (at home) |
| Introducing researcher terminology | The interviewer might use personal terms when asking clarifying questions (i.e. "did you relate here to providing an overview") | There is a standard procedure in which the respondent is asked to compare their features to a list of pre-defined features |
| Researcher Intervention in the clarification process | The interviewer intervenes to clarify meaning | The clarification takes place within the peer discussion and pre-defined feature list |
| Resulting data | Transcript of verbal discussion | Compact written text |
| Constraints on respondent answer | Conversation, constrained by interviewer questions | Participants must differentiate between features, preferences and reasons |
| Separating intervention with the interviewer from the data | Interaction with the interviewer via. clarifying questions (i.e. what do you mean, can you please elaborate) is an integral part of the data | The pre and post individual worksheet can be separated from the interaction with peers where respondents clarify their thoughts |

Yet, there are also cons to the GAIQ tool, such as the inability to respond on the spot to better clarify respondent's ideas, or the conciseness of written responses relative to spoken ones. Researchers should carefully consider their goals and resources before choosing either of these data collection sources.